\documentclass{emulateapj}

\usepackage{amsmath}
\usepackage{graphicx}
\usepackage{natbib}

\newcommand{\hii}{H~{\sc ii}}

\newcommand{\kb}{k_\mathrm{B}}
\newcommand{\tmin}{T_\mathrm{min}}

\newcommand{\mmax}{M_\mathrm{max}}
\newcommand{\mdisk}{M_\mathrm{disk}}
\newcommand{\msinks}{M_\mathrm{sinks}}
\newcommand{\mtot}{M_\mathrm{tot}}
\newcommand{\nsinks}{N_\mathrm{sinks}}

\begin{document}

\title{Limiting Accretion onto Massive Stars by
   Fragmentation-Induced Starvation}

\author{Thomas Peters\altaffilmark{1,2}}
\email{thomas.peters@ita.uni-heidelberg.de}

\author{Ralf S. Klessen\altaffilmark{2,3}, Mordecai-Mark Mac Low\altaffilmark{4}, Robi
  Banerjee\altaffilmark{2}}

\altaffiltext{1}{Fellow of the Baden-W\"{u}rttemberg Stiftung}
\altaffiltext{2}{Zentrum f\"{u}r Astronomie der Universit\"{a}t Heidelberg,
Institut f\"{u}r Theoretische Astrophysik, Albert-Ueberle-Str. 2,
D-69120 Heidelberg, Germany}
\altaffiltext{3}{Kavli Institute for Particle Astrophysics and Cosmology, 
Stanford University, Menlo Park, CA 94025, U.S.A.}
\altaffiltext{4}{Department of Astrophysics, American Museum of Natural History, 
79th Street at Central Park West, New York, New York 10024-5192, USA}

\begin{abstract}
  Massive stars influence their surroundings through
  radiation, winds, and supernova explosions far out of proportion to
  their small numbers. 
          However, the physical processes
  that initiate and govern the birth of massive stars
          remain poorly
  understood.
  Two widely discussed models are monolithic collapse of molecular cloud
  cores and competitive accretion. To learn more about massive star formation,
  we perform and analyze simulations of the collapse of rotating, massive, cloud cores
  including radiative heating by both non-ionizing and ionizing radiation
  using the FLASH adaptive mesh refinement code. These simulations show
             fragmentation from gravitational instability in the
             enormously dense accretion flows required to build up
             massive stars.
  Secondary stars form rapidly in these
  flows and accrete mass that would have otherwise been consumed by the massive
  star in the center, in a process that we term fragmentation-induced starvation.
  This explains why massive stars are usually found as members of high-order
  stellar
          systems that themselves belong to
  large clusters
          containing stars of all masses.
  The radiative heating does not prevent fragmentation,
  but does lead to a higher Jeans mass, resulting in fewer and more massive stars
  than would form without the heating. This mechanism reproduces the observed
  relation between the total stellar  mass in the cluster and the mass of the
  largest star. It predicts strong clumping and filamentary structure in the
  center of collapsing cores, as has recently been observed. We speculate that
  a similar mechanism will act during primordial star formation.
\end{abstract}

\maketitle

\section{Introduction}
\label{sec:intro}
Understanding the formation of massive stars is of pivotal importance in modern astrophysics.
Massive stars are rare and short lived. However, they are also very bright and allow us to
reach out to the far ends of the universe. For example, the most distant galaxies in the
Hubble Ultra Deep Field \citep{beckwithetal06} are all characterized by vigorous high-mass star formation.
Understanding the origin of massive stars, at present and at early times, therefore is a
prerequisite to understanding cosmic history. In our own Milky Way, high-mass stars
contribute the bulk of the UV radiation field. This radiation can ionize hydrogen and dissolve
molecules. Expanding \hii\ regions, bubbles of ionized gas surrounding massive stars, have been
identified as an important source  of interstellar turbulence. The complex interplay between
gas dynamics and radiation thus is a key element of the evolution of the interstellar medium
(ISM) and the Galactic matter cycle. Furthermore, stars are  the primary source of chemical
elements heavier than the hydrogen, helium, and lithium that were produced in the Big Bang.
Again, massive stars have a disproportionally large share in the enrichment history of the
Galaxy and the universe as a whole. It is the violent supernova explosions associated with
the death of high-mass stars that contribute most of the metals. Because they insert large
amounts of energy and momentum input, supernovae at the same time strongly stir up the ISM
and provide for the effective mixing of the newly bread elements.

Despite their importance, the physical processes that initiate and control the build-up of
massive stars are still not well understood and subject of intense debate
\citep{maclow04,zinnyork07,mckee07}. Because their
formation time is short, of order of $10^5\,$yr, and because they form deeply embedded in
massive cloud cores, very little is known about the initial and environmental conditions of
high-mass stellar birth. In general high-mass star forming regions are characterized by more
extreme physical conditions than their low-mass counterparts, containing cores of size, mass,
and velocity dispersion roughly an order of magnitude larger than those of cores in low-mass
regions \citep[e.g.][]{jijmeyad99,garliz99,kurtzetal00,beutheretal07,motteetal08}.
Typical sizes of cluster-forming clumps are $\sim 1\,$pc,
they have mean densities of $n \sim 10^5$ cm$^{-3}$, masses of $\sim 10^3\,$M$_\odot$ and
above, and velocity dispersions ranging between $1.5$ and $4 \,$km$\,$s$^{-1}$. Whenever
observed with high resolution, these clumps break up in even denser cores, that are believed
to be the immediate precursors of single or gravitationally bound multiple massive protostars. 

Massive stars usually form as members of multiple stellar systems \citep{hohaschik81,lada06,zinnyork07}
which themselves are parts of larger clusters \citep{ladalada03,dewitetal04,testietal97}. This
fact
   adds
additional challenges
   to
the interpretation of observational data from high-mass
star forming regions as it is difficult to disentangle mutual dynamical interactions from the
influence of individual stars \citep[e.g.][]{gotoetal06,linzetal05}.
Furthermore, high-mass stars
  reach
the main sequence while still accreting. Their Kelvin-Helmholtz
pre-main sequence contraction
   time
is considerably shorter than
   their accretion time.
Once
  a
star has reached a mass of about $10\,$M$_\odot$ its spectrum becomes UV dominated and it begins
to ionize its environment. This means that accretion as well as ionizing and non-ionizing radiation
needs to be considered in concert \citep{keto02b,keto03,keto07,petersetal10a,petersetal10b}. It has been realized
decades ago that in simple 1-dimensional collapse models the outward radiation force on the accreting
material should be significantly stronger than the inward pull of
gravity \citep{larsstarr71,kahn74,wolfcas87}
in particular when taking dust opacities into account.
   Since
stars as massive as $100 - 150\,$M$_\odot$ have
been observed \citep{bonanosetal04,figer05,rauwetal05} a simple
         spherically
symmetric approach to
high-mass star formation is doomed to fail. 

Consequently, two different models for massive star formation have been proposed. The first one takes
advantage of the fact that high-mass stars always form as members of stellar clusters. If the central
density in the cluster is high enough, there is a
         chance
that low-mass protostars collide and so
successively build up more massive objects \citep{bonbatezin98}. As the radii of protostars
usually are considerably larger than the radii of main sequence stars in  the same mass range this
could be a viable option. However, the stellar densities required to produce very massive stars are
still extremely high and seem inconsistent with the observed values of Galactic star clusters 
\cite[e.g.][and references therein]{pozwetal10}.
          An
alternative approach is to argue that high-mass stars form just like low-mass stars by accretion
of ambient gas that goes through a rotationally supported disk caused by angular momentum conservation
\citep{mckeetan03}.
Indeed such disk structures are observed around a number of high-mass protostars
\citep{chini04,chini06,jiangetal08,daviesetal10}. Their presence
breaks any spherical symmetry that might have been present in the initial cloud and thus solves the
opacity problem. Radiation tends to break out along the polar axis, while matter is transported inwards
through parts of the equatorial plane shielded by the disk. Hydrodynamic simulations in two
\citep{yorke02} and three (\citealt{krumholzetal09}, \citealt{kuiperetal10}, in prep.)
dimensions using a flux-limited diffusion approach to the transport of non-ionizing radiation strongly
support this picture. \citet{krumholzetal09} find non-axisymmetric Rayleigh-Taylor instabilities
using gray radiation transfer, whereas \citet{kuiperetal10}, using frequency-dependent radiation
transfer, do not find such instabilities, but nevertheless find strong disk accretion.
Strong accretion continues when ionizing radiation is included using ray-tracing methods in
three dimensions
\citep{daleetal05,daleetal07b,petersetal10a,petersetal10b}.
In a clustered environment or in individual collapsing cores where the
disk becomes
         gravitationally
unstable,
the three-dimensional models show that
material flows along dense, opaque filaments whereas the radiation
escapes through optically thin channels
of low-density material.
It has been demonstrated that even
ionized material can be accreted, if the accretion flow is strong enough. \hii\ regions are gravitationally
trapped at that stage, but soon begin to rapidly fluctuate between trapped and extended states, in
agreement with observations \citep{petersetal10a,galvmadetal10inprep}. Over time, the same
ultracompact \hii\ region can expand anisotropically, contract again, and take on any of the observed
morphological classes \citep{woodchurch89,kurtzetal94,petersetal10a,petersetal10b}. In their
extended phases, expanding \hii\  regions drive bipolar neutral outflows characteristic of high-mass
star formation \citep{petersetal10a}. 

Another key fact that any theory of massive star formation must account for is the apparent presence
of an upper mass limit. No star more massive than  $100 - 150\,$M$_\odot$ has been observed \citep{massey03}.
This holds for the Galactic field, however, it is also true for young star clusters
that are massive enough so that purely random sampling of the initial mass function (IMF)
\citep{kroupa02,chabrier03} without upper mass limit should have yielded stars above
$150\,$M$_\odot$ (\citealt{weidkroup04,figer05,oeycla05,weidetal10}, see however, \citealt{selmel08}).
This immediately raises the question
of what is the physical origin of this apparent mass limit. It has been speculated before that
      radiative
stellar feedback might be responsible for this limit \citep[see, e.g.,][]{zinnyork07} or
alternatively that the internal stability limit of stars with non-zero metallicity lies in this
mass regime \citep{appen70a,appen70b,appen87,baraetal01}.
However,
    fragmentation could also
limit protostellar mass growth. Indeed, this is what
we see
    in
the simulations discussed here. The likelihood of fragmentation to occur and the number of fragments
to form depends sensitively on the physical conditions in the star-forming cloud and its initial and
environmental parameters \citep[see, e.g.,][]{krumetal10,girietal10}. Understanding the build-up of massive
stars, therefore, requires detailed knowledge about the physical processes that initiate and regulate the
formation and dynamical evolution of the molecular clouds these stars form in 
\citep{vazsemetal09}.

We argue that ionizing radiation, just like its non-ionizing, lower-energy counterpart,
         cannot
shut off the accretion flow onto massive stars. Instead it is the dynamical processes
in the gravitationally unstable accretion
         flow
that inevitably
      occurs
during the collapse of
high-mass cloud cores that
         control
the mass growth of individual protostars.  Accretion onto the
central star is shut off by the fragmentation of the disk and the formation of  lower-mass companions
that intercept inward moving material. We call this process fragmentation-induced starvation and
argue that it occurs unavoidably in regions of high-mass star formation where the mass flow onto the
disk exceeds the inward transport of matter due to viscosity only and thus renders the disk unstable
to fragmentation.
We speculate that
        fragmentation-induced starvation
is important not only for present-day star formation
but also in the primordial universe during the formation of metal-free Population III stars.
Consequently, we expect these stars to be in binary or small number multiple systems and to be of
lower mass than usually inferred \citep{abeletal02,brommetal09}. Indeed, current
numerical simulations provide the first hints that this might be the case
\citep{clarketal08,turketal09,stacyetal10}.

     In the current study, we analyze the simulations by
     \citet{petersetal10a} with special focus on the mass growth
     history of the individual stars forming and the physical
     processes that influence their accretion rate.
     We
briefly review the  numerical method we use and the assumptions and approximations behind it in
Section \ref{sec:method}. Then we describe our findings in Section \ref{sec:results} and discuss them
in the context of present-day and primordial star formation in Section
\ref{sec:discussion}. We summarize
and conclude in Section \ref{sec:summary}.

\section{Method and Assumptions}
\label{sec:method}

We present three-dimensional, radiation-hydrodynamical simulations of massive star formation that include
heating by ionizing and non-ionizing radiation using the adaptive-mesh code FLASH \citep{fryxell00}.
We use our improved version of the hybrid-characteristics raytracing method \citep{rijk06,petersetal10a}
to propagate the radiation on the grid and couple sink particles \citep{federrathetal10}, which we use as
models of protostars, to the radiation module via a prestellar model \citep{petersetal10a}.

The simulations start with a $1000\,$M$_\odot$ molecular cloud. The cloud has a constant density core
of $\rho = 1.27 \times 10^{-20}\,$ g\,cm$^{-3}$ within a radius of $r = 0.5\,$pc and then falls off
as $r^{-3 / 2}$ until $r = 1.6\,$pc. The initial temperature of the cloud is $T = 30\,$K. The whole
cloud is set up in solid body rotation with an angular velocity $\omega = 1.5 \times 10^{-14}\,$s$^{-1}$ 
corresponding to a ratio of rotational to gravitational energy $\beta = 0.05$ and a mean specific angular
momentum of $j = 1.27 \times 10^{23}\,$cm$^2$s$^{-1}$.

We follow the gravitational collapse of the molecular cloud with the adaptive mesh until we reach a
cell size of $98\,$AU. We create sink particles at a cut-off density of
$\rho_\mathrm{crit} = 7 \times 10^{-16}\,$g\,cm$^{-3}$. All gas within the accretion radius of
$r_\mathrm{sink} = 590\,$AU  above $\rho_\mathrm{crit}$ is accreted to the sink particle
if it is gravitationally bound to it. The Jeans mass on the highest refinement level is
$M_\mathrm{jeans} = 0.13$\,M$_\odot$.

The adaptive mesh technique allows us
     to
resolve the gravitational collapse of the gas from the parsec scale
down to a few hundred AU. At higher spatial resolution of only several ten AU, the gas becomes optically
thick to non-ionizing radiation, and scattering effects must be taken into account. Since our
cut-off density is more than two orders of magnitude smaller than the onset of the optically thick
regime at $\sim 10^{-13}$\,g\,cm$^{-3}$~\citep[e.g.][]{larson69} and we are focussing in our analysis
on large-scale effects on the stellar cluster scale, we expect the raytracing approximation to be valid.
Feedback by radiation pressure, which plays a role on the very small scales and is neglected
in our model, is dynamically unimportant on these large scales \citep{krummatzner09}.

We discuss three simulations (see Table~\ref{tab:colsim}). In the first simulation (run~A),
we only allow 
for
the formation of a single sink particle and suppress the formation of secondary
sink particles artificially by introducing the dynamical temperature floor
\begin{equation}
\tmin = \frac{G \mu}{\pi \kb} \rho (n \Delta x)^2
\end{equation}
with Newton's constant $G$, mean molecular weight $\mu$, Boltzmann's constant $\kb$, local gas
density $\rho$, and cell size $\Delta x$. The temperature floor ensures the sufficient
resolution of the Jeans length with $n \geq 4$ cells, which avoids artificial fragmentation
\citep{truelove97}. The suppression of secondary sink formation guarantees that the accretion flow
around the massive star is not weakened by the fragmentation of the disk, which would otherwise
inevitably lead to the formation of companion stars
          that
limit the growth of the massive stars
in the cluster (see Section~\ref{sec:results}). In the second simulation (run~B), this
dynamical temperature floor is not applied, and a whole stellar cluster forms during the
  course of the simulation.
The third simulation (run~D) is a control run that allows us to study the
influence of radiation feedback on the stellar cluster evolution. As in run~B, a
    small
stellar cluster forms, but the stars emit 
   neither ionizing nor non-ionizing radiation.

To demonstrate convergence of our study, we have also run an
additional simulation 
  at twice the numerical resolution. This simulation, run~D+, which we
  ran for 0.66~Myr, is identical
to run~D, except that the cell size at
highest grid refinement is $49\,$AU, the sink particle radius is $r_\mathrm{sink} = 312\,$AU and the
sink particle cut-off density is $\rho_\mathrm{crit} = 2.5 \times 10^{-15}\,$g\,cm$^{-3}$.
Since 
           we ran it for only a fraction of the time covered by 
the other simulations, we will
not discuss it at length, but the 
           model ran for long enough 
to show that the
relation between total stellar cluster mass and the mass of the largest star is the same is in
the lower resolution simulations (see Section~\ref{subsec:compaccr} for further discussion).

The numerical method along with its inherent physical limitations is discussed in detail
in \citet{petersetal10a}.

\section{Results}
\label{sec:results}

\begin{table*}
\begin{centering}
\caption{Overview of collapse simulations. \label{tab:colsim}}
\begin{tabular}{ccccccc}
\tableline
\tableline
Name & Resolution & Radiative Feedback & Multiple Sinks &
$\msinks$\,(M$_\odot$) & $\nsinks$ & $\mmax$\,(M$_\odot$) \\
\tableline
Run A & 98~AU & yes & no  & 72.13  &  1 & 72.13 \\
Run B & 98~AU & yes & yes & 125.56 & 25 & 23.39 \\
Run D & 98~AU & no  & yes & 151.43 & 37 & 14.64 \\
\tableline
\end{tabular}
\tablecomments{We have run a simulation with a single sink particle (run~A) to
probe the upper mass limit with radiative feedback as well as simulations with
multiple sink particles to study stellar cluster formation with (run~B) and without
(run~D) radiative feedback. The simulations differ largely in the total mass in sink
particles ($\msinks$), the total number of sink particles ($\nsinks$) and the mass of the
most massive star ($\mmax$) at the end of the simulation.
           We
follow the notation
introduced in \citet{petersetal10a}, where additional low-resolution runs~Ca und Cb were discussed.}
\end{centering}
\end{table*}

\begin{figure}
\centerline{\includegraphics[width=8cm]{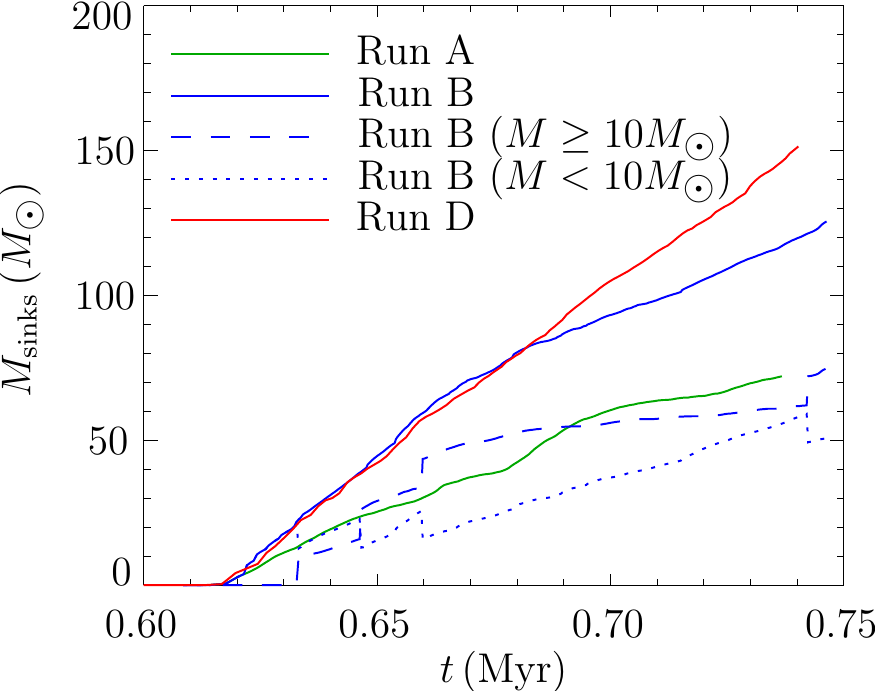}}
\centerline{\includegraphics[width=8cm]{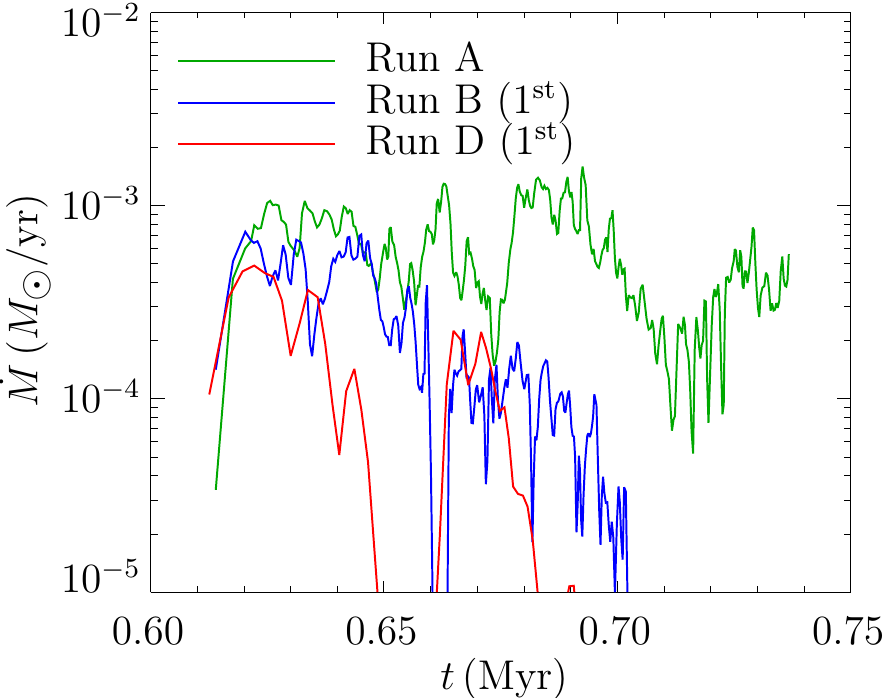}}
\caption{ {\em top} 
Total accretion history of all sink particles combined forming in
runs A, B, and D. While the heating by non-ionizing radiation does not affect the
total star formation rate, the ionizing radiation appreciably reduces
the total rate at which gas converts into stars once the most massive object has stopped accreting
and its \hii\ region can freely expand. This effect is not compensated by triggered star formation
at the ionization front, which is never observed during the simulation.
The slope of the total accretion history in run~B goes down because the massive stars (dashed line)
accrete at a decreased rate, while the low-mass stars (dotted line) keep accreting at the same rate.
{\em bottom} Instantaneous accretion rate as function of time of the first sink
particle to form in the three runs. These are generally
the most massive sink particles during most of the simulation. 
While
the accretion onto the star in run~A never stops, the massive stars in run~B and~D
are finally starved of material. Since the radiative heating suppresses fragmentation
in run~B, the final mass of the star is almost twice as high as in run~D.}
\label{fig:accretion}
\end{figure}

\subsection{Accretion History}

In this section we compare the protostellar mass growth rates from
   our three runs, with a single sink particle (run~A), multiple sinks
   and radiative heating (run~B), and multiple sinks with no radiative
   heating (run~D).
As already discussed by \citet{petersetal10a},
      when
only the central sink particle
   is
allowed to form (run~A), nothing
stops the accretion flow to the center. Figure~\ref{fig:accretion} shows that the central
protostar grows at a rate $\dot{M} \approx 5.9 \times 10^{-4}\,$M$_{\odot}\,$yr$^{-1}$ until
we stop the calculation when the star has reached $72\,$M$_{\odot}$. The growing star ionizes
the surrounding gas, raising it to high pressure. However this hot bubble soon breaks out above
and below the disk plane,
   without affecting
the gas flow in the
         disk\footnote{We will hereafter refer to the flattened,
           dense, accretion flow that forms in the midplane of our
           rotating core as a disk.  However, it is not necessarily a
           true Keplerian, viscous, accretion disk, which probably
           only forms within the central few hundred astronomical
           units, unresolved by our models.}
midplane much. In
particular, it cannot halt the accretion onto the central star. Similar findings have also been
reported from simulations focussing on the effects of non-ionizing radiation acting on somewhat
smaller scales \citep{yorke02,krumkleinmckee07,krumholzetal09,sigalottietal09}.
Radiation pressure is not able to stop accretion onto massive stars and is dynamically unimportant,
except maybe in the centers of dense star clusters near the Galactic center \citep{krummatzner09}. 

The situation is different when
the disk can fragment and form multiple sink particles.  Initially the
mass growth of the central protostar in runs~B and D is comparable to
the one in run A. As soon as further protostars form in the
gravitationally unstable disk, they begin to compete with the central
object for accretion of disk material.  However, unlike in the
classical competitive accretion picture \citep{bonnell01a,bonetal04},
it is not the most massive object that dominates and grows
disproportionately fast.  Figure~\ref{fig:accretion} shows that,
although the accretion rates of the most massive stars ($M \geq
10\,$M$_\odot$) steadily decrease, the low-mass stars ($M <
10\,$M$_\odot$), keep accreting at the same rate.  Although the
detailed mass distribution of the low-mass stars depends on numerical
resolution \citep{federrathetal10}, the net effect of their accretion
should not, so long as they can form at all.  The successive formation
of low-mass objects in the disk at increasing radii limits subsequent
growth of the more massive objects in the inner disk. Material that
moves inwards through the disk accretes preferentially onto the sinks
at larger radii.

The key to the process, as already discussed
by \citet{bate00}, is that fragments that form later at larger radii will
preferentially accrete larger-angular momentum material.  This will
happen on a much shorter time than the timescales required to
redistribute this angular momentum to other parts of the disk at
larger radii by viscous or gravitational torques. Similar effects are
discussed in the disk fragmentation studies by \citet{kratteretal10}.

This behavior is found in models of 
low-mass 
 protobinary disks, where again the secondary accretes at a higher rate than the primary. Its orbit
around the common center of gravity scans larger radii and hence it encounters material that moves
inwards through the disk before the primary star. This drives the system towards equal masses and
circular orbits \citep{bateetal97,bate00}.
In our
simulations, after a certain transition period hardly any gas makes it all the way to the center and
the accretion rate of the first sink particle drops to almost zero. This is the essence of the
fragmentation-induced starvation process. In run~B, it prevents any
star
         from
reaching a mass
larger than $25\,$M$_\odot$. The Jeans mass in run~D is smaller than in run~B because of the
lack of accretion heating, and consequently the highest mass star in run~D grows to less than
$15\,$M$_\odot$.
In comparison, \citet{krumkleinmckee07} described simulations starting with
a ten times less massive core than ours, using both an isothermal equation
of state and including accretion heating.  In both cases, objects with
roughly half the mass of our most massive object formed.  The isothermal
case formed an only slightly less massive object than the one including
accretion heating, just as in our models.  The smaller final masses of their
objects compared to ours seems mainly just to be a result of their smaller
initial core masses.

Inspection of Figure~\ref{fig:accretion} reveals additional aspects of the process. We see that
the total mass of the sink particle system increases at a faster rate in the multiple sink
simulations, run~B and D, than in the single sink case, run A. This is understandable, because
as more and more gas falls onto the disk it
         becomes
more and more unstable
         to fragmentation, so
as time goes by additional sink particles form at larger
and larger radii.
Star formation occurs
in a larger volume
         of the
disk, and mass growth is not limited by the disk's ability to
transport matter to
         its center by gravitational or viscous
torques
(compare Section~\ref{sec:diskinst}).
As a result the overall star-formation
rate is larger than in run~A.

Since the accretion heating raises the Jeans
    mass and
length in run~B,
the total number of sink particles is higher in run~D than in run~B,
      and the
stars in run~D generally reach a lower mass than in
run~B (compare Section~\ref{subsec:compaccr}). These two effects
cancel out
to lead to the same overall star formation
          rate for some time.
At one point in the evolution, however, also
the total accretion rate of run~B drops below that of run~D.  At time
$t \approx 0.68\,$Myr the accretion flow around the most massive star
has attenuated below the value required to trap the \hii\ region. It
is able to break out and affect a significant fraction of the disk
area. A comparison with the mass growth of run~D clearly shows that
there is still enough gas available to continue constant cluster
growth for another $50\,$kyr or longer, but the gas can no longer
collapse in run~B.  Instead, it is swept up in a shell surrounding the
expanding \hii\ region. 

It is also notable that the expanding ionization front around the most
massive stars does not trigger any secondary star formation,
which suggests that triggered star formation \citep{elmelad77} may not be as efficient as expected,
at least on the scales considered here.

\begin{figure*}
\centerline{\includegraphics[width=8cm]{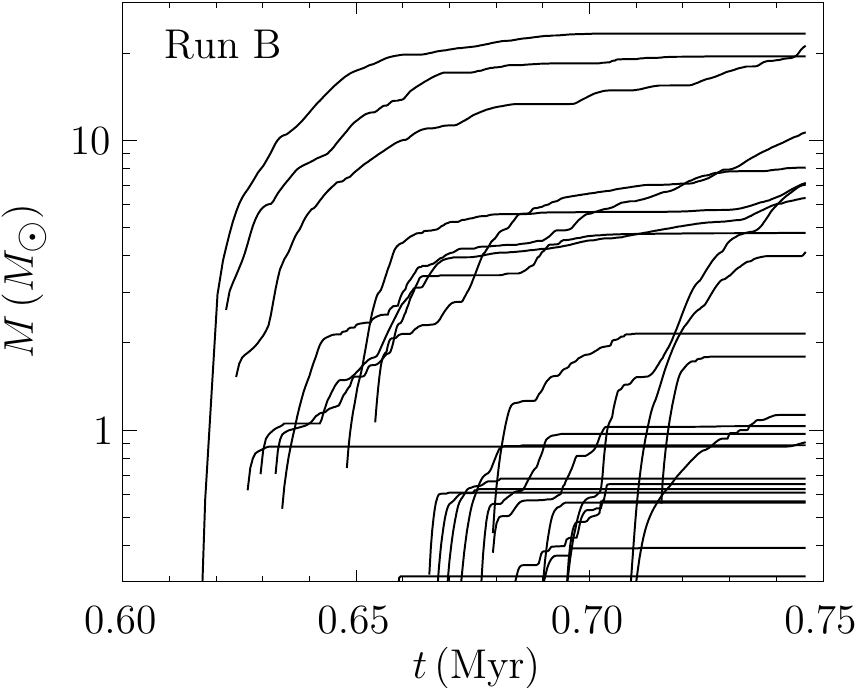}
\includegraphics[width=8cm]{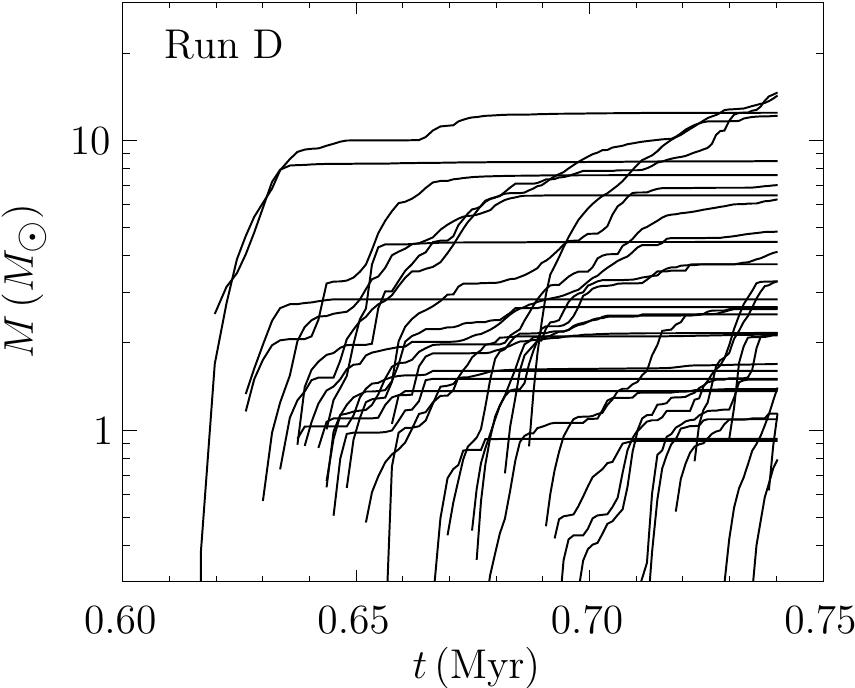}}
\caption{Individual accretion histories for run~B and run~D. The figure shows the stellar
masses as function of time for all sink particles that form in run~B ({\em left}) and run~D ({\em right}).
Because the Jeans mass is lower without radiative feedback (run~D),
         many
more sink particles form in run~D than
in run~B, and the mass of the most massive stars is also lower.}
\label{fig:history}
\end{figure*}

Figure~\ref{fig:history} shows the individual accretion histories of each of the sink
particles in run~B and run~D. Radiative heating cannot prevent disk fragmentation but
raises the local Jeans mass. Hence, as discussed above, much fewer stars form in run~B than
in run~D, and the mass of the most massive stars in run~B is higher. It is also evident
from the figure that star formation is much more intermittent in the case with radiation feedback
(run~B). The reason for
this behavior is that the star formation process is controlled by the local Jeans mass,
which depends to a large degree on how the filaments in the disk shield the radiation
(compare Section~\ref{sec:diskinst}),
            as
  previously noted by \citet{daleetal05,daleetal07b} and \citet{krumholzetal09}.
Shielding can lower the Jeans mass temporarily
and thereby allow gravitational collapse that would not have occurred otherwise.

The accretion histories also reveal large differences between sink particles within the same
simulation. For example, in run~D a sink particle that forms around $t \approx 0.688\,$Myr accretes
very rapidly and becomes one of the most massive stars in the cluster while other stars only
accrete sporadically, interrupted with long intervals of no accretion at all, and some others
even accrete only shortly after their formation. There are several reasons for these
differences. First, accretion can stop when stars are ejected from the cluster by
$N$-body interactions. This happens mostly to low-mass stars with masses $M \lesssim 2$M$_\odot$.
Second, intermediate-mass stars can be starved
         of
material by the formation of companions.
This is the essential element of the fragmentation-induced starvation
scenario. A star
          that
can grow to intermediate masses is embedded in an accretion flow from a larger gas reservoir.
If this reservoir is gravitationally unstable, it can
         collapse to form
companions. This may explain why the intermediate-mass stars in run~B
seem to have shorter time intervals
of no accretion. Since the Jeans mass is higher in this case, the formation of companions
is suppressed, and this allows the star to accrete for longer times. Third, the stars
are not fixed at a position in the disk plane but
         move quickly
within the disk by gravitational interaction with their surrounding dense gas and with nearby stars.
Hence, they are not tied to
the filament they form in but can move away from it. If the star moves into a low-density
void in the disk, accretion will stop until the star is embedded in higher density gas again.
The sink particle in run~D mentioned above accretes so vigorously because it moves 
along
with its parental filament over a long time and no companions form around it. Fourth,
accretion can stop when the star resides within a low-density \hii\ region. This applies
to the first two massive stars in run~B. Their accretion flow had become so weak that the
ionizing radiation was able to isolate the stars from the high-density gas in the disk
and stopped accretion.

\subsection{Disk Growth and Disk Instability}
\label{sec:diskinst}

\begin{figure}
\centerline{\includegraphics[width=8cm]{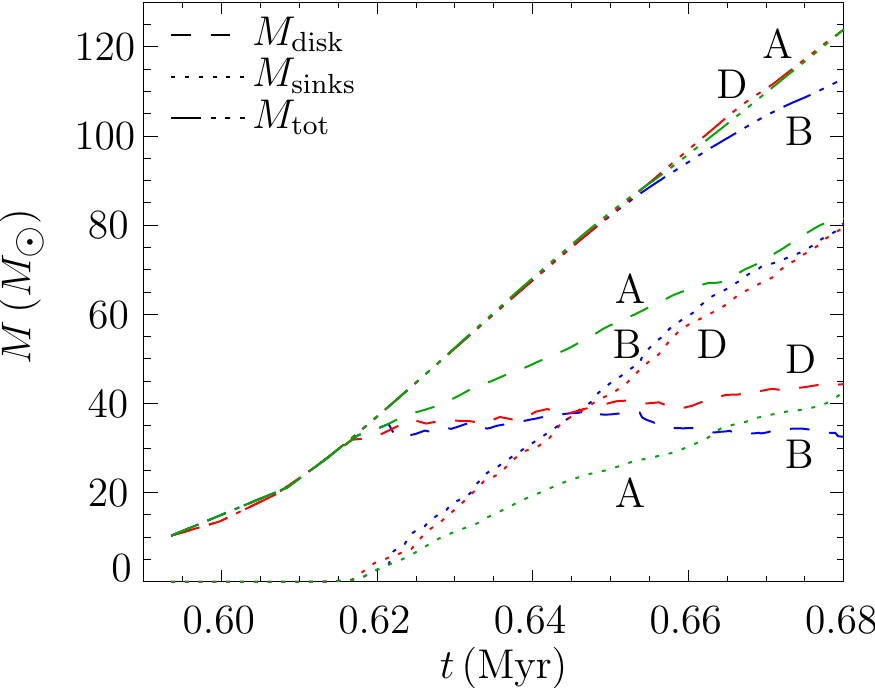}}
\caption{Disk growth and sink particle formation. The plot shows the mass $\mdisk$ 
of non-accreted gas contained
in a control volume around the disk, the mass $\msinks$ of all sink particles and the total mass
$\mtot = \mdisk + \msinks$ for run~A (green), run~B (blue) and run~D (red). The disk mass
in runs~B and D is kept almost constant by subsequent sink particle formation, while the disk
in run~A continuously grows. The deviation between the disk masses in runs~B and D 
at late times
is caused
by ionization-driven outflows, but these do not affect the total star formation rate.}
\label{fig:massinvolume}
\end{figure}

To quantify the accretion of mass onto the disk and the subsequent instability, we calculate
the amount of mass in a control volume that encloses the disk at all times. This control
volume is the same for all runs and has dimensions $0.24\,$pc$\times 0.24\,$pc$\times 0.015\,$pc.
We show in Figure~\ref{fig:massinvolume} the mass $\mdisk$ 
of non-accreted gas 
contained in the control volume,
the mass $\msinks$ contained in all sink particles as well as the total mass $\mtot = \mdisk
+ \msinks$ for runs~A, B and D. The disk mass in run~A is much larger than in the multiple
sink runs B and D since the mass cannot be absorbed by secondary sink particles. Instead, the mass
accumulates in the disk plane. Consequently, the star in run~A is embedded in a strong accretion
flow at all times, which facilitates accretion at an approximately constant rate despite radiation
feedback. The evolution in runs~B and D differs from the mass growth in run~A as soon as 
secondary
sink particles form. 
The continuous formation of new sink particles in an expanding region around the central star in
these runs keeps $\mdisk$ almost constant. Figure~\ref{fig:massinvolume} shows that at
$t \approx 0.64\,$Myr, outflow of material from the central \hii\ region
\citep{petersetal10a}
reduces $\mdisk$ even further in run~B, so that it falls slightly below the case without
feedback (run~D) at late times.  However, accretion onto secondary sink particles had
already cleared the central region, so that this is a relatively minor effect.  The
accretion onto sink particles is not affected by the central ionization and stays at a
constant rate until the \hii\ region can break free (compare Figure~\ref{fig:accretion}).

\begin{figure*}
\centerline{\includegraphics[width=450pt]{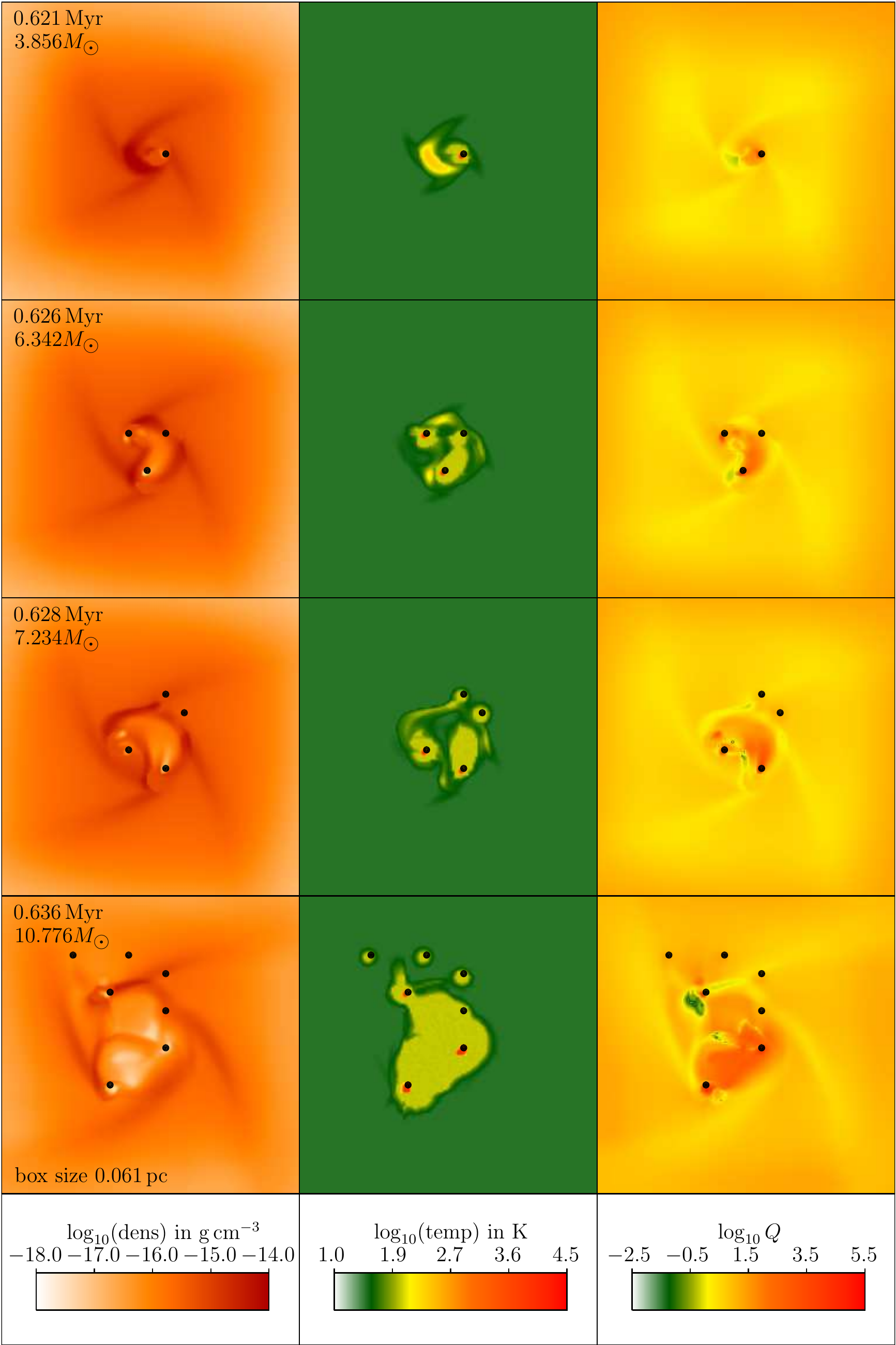}}
\caption{Early phase of disk instability in run~B. The panels show slices of density, temperature and
the Toomre $Q$-parameter at four different times. Each frame shows the
simulation 
  time 
and the
mass of the most massive star in the cluster. The black dots indicate the positions of sink particles.
One can see how the stellar radiation initially heats up the disk locally, which enhances the stability.
The dense filaments, however, shield the radiation, and the cold material within and behind the filaments
becomes unstable again.}
\label{fig:toomreq}
\end{figure*}

We investigate the initial instability of the disk with an analysis of the Toomre $Q$-parameter
\begin{equation*}
Q = \left| \frac{2 c_\mathrm{s} \Omega}{\pi \Sigma G} \right|
\end{equation*}
with
sound speed $c_\mathrm{s}$,
angular velocity $\Omega$,
surface density $\Sigma$, and Newton's constant $G$. The disk is
linearly stable for $Q > 1$ and linearly unstable
for $Q < 1$ \citep{toomre64,goldlynd65}. The initial phases of disk instability in run~B are displayed
in Figure~\ref{fig:toomreq}. The figure shows slices of density, temperature and $Q$ in
the disk plane for four different times. One can clearly see that the most unstable parts of
the disk are the filamentary structures that form 
as the disk grows in mass. The heating
by stellar radiation can suppress instability locally, but shielding by the dense filaments
prevents the whole disk from becoming stable and restricts the heating to small regions near the
center of the disk
           that
are surrounded by filaments. This shielding makes it possible for star
formation to progress radially outwards despite accretion heating by the stars. 
The disk remains sufficiently cool at the inner edge for gravitational instability to set in and  star formation  proceeds
inside-out in the disk plane. Hence, the effect of the filamentary structures in the disk is twofold:
they are so dense that they render the disk unstable locally; and, because of
their high density, they can effectively shield the radiation from the stellar cluster near the center
of the disk, so that
         radiative heating does not stabilize the outer parts of the disk.
Indeed, high-resolution observations of high-mass accretion disk candidates
\citep{beutetal09} provide some evidence for fragmentation and the presence of
substructure on $\sim 1000\,$AU scales (as also proposed by \citet{krumetal07}).

\begin{figure}
\centerline{\includegraphics[width=167pt]{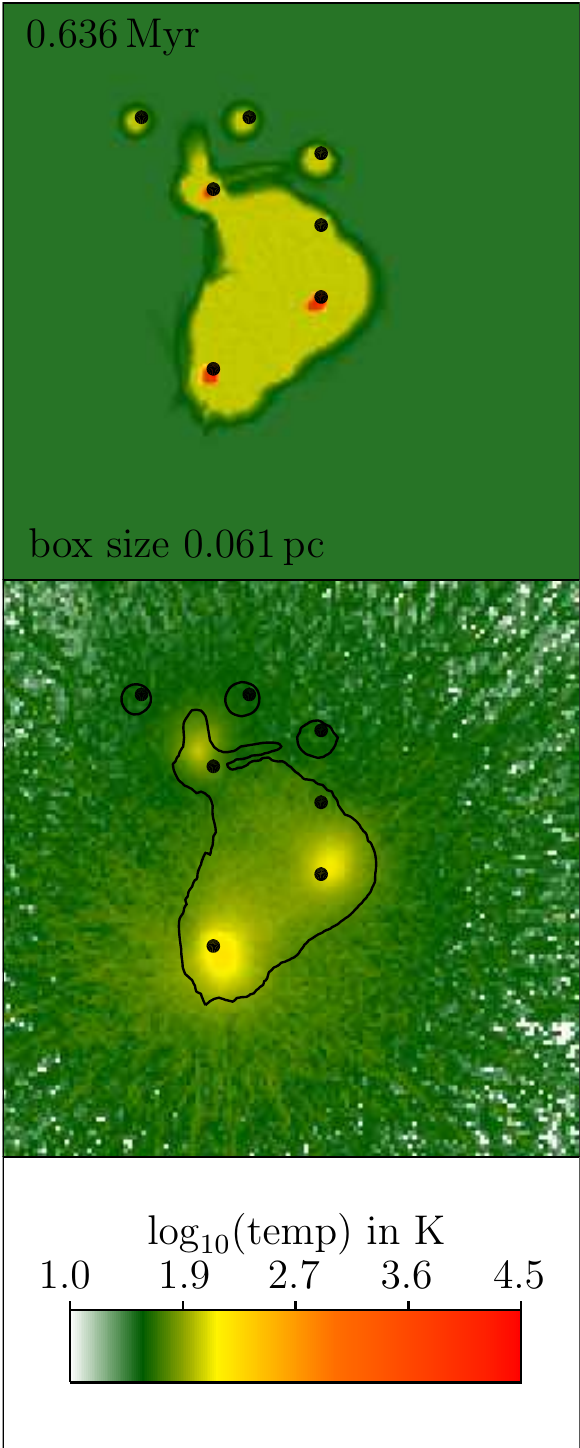}}
\caption{Comparison of heating by direct and diffuse radiation. The
  slices show the gas temperature from the simulation snapshot 
   in the last row of Figure~\ref{fig:toomreq} ({\em
    top}) and the Monte Carlo dust temperature generated by RADMC-3D
  for the same snapshot
  ({\em bottom}). The dust temperature from diffuse radiation lies
  below the gas temperature almost everywhere outside the region
  heated by direct radiation ({\em contour}). The
  black dots indicate the positions of sink particles.}
\label{fig:diffuseplot}
\end{figure}

Since the filaments that shield the outer parts of the disk from
radiation are optically thick, with an optical depth for non-ionizing
radiation of several 
  tens,
it is important to estimate the degree
to which our simulations, based on a raytracing technique to
propagate the radiation on the grid, are affected by the lack of heating by diffuse
radiation. To test this, we use the adaptive-mesh radiative transfer
program
RADMC-3D\footnote{http://www.mpia.de/homes/dullemon/radtrans/radmc-3d/index.html}. RADMC-3D
is based on the standard Monte Carlo method of \citet{bjorkmanwood01} in
combination with Lucy's method of treating optically thin regions
\citep{lucy99}. It is the successor of the RADMC code
\citep{dullemdom04} and has been used previously to generate maps of
dust emission 
  from these simulations
\citep{petersetal10b}.  We have calculated
self-consistent dust temperatures of the simulation snapshot shown in
the last row of Figure~\ref{fig:toomreq}. Assuming that dust and gas
temperatures in the disk plane are equal, we can compare the Monte
Carlo dust temperature with the simulation gas temperature, 
  as shown in Figure
\ref{fig:diffuseplot}.  The comparison demonstrates that direct
heating dominates over diffuse radiation in regions more than
$\sim500$~AU away from the stars. 
  The diffusely heated regions lie completely within the region heated
  in any case by direct radiation, so our 
raytracing method accurately
describes the shielding by the filaments of the cold disk region where
secondary fragmentation proceeds.

\begin{figure}
\centerline{\includegraphics[width=8cm]{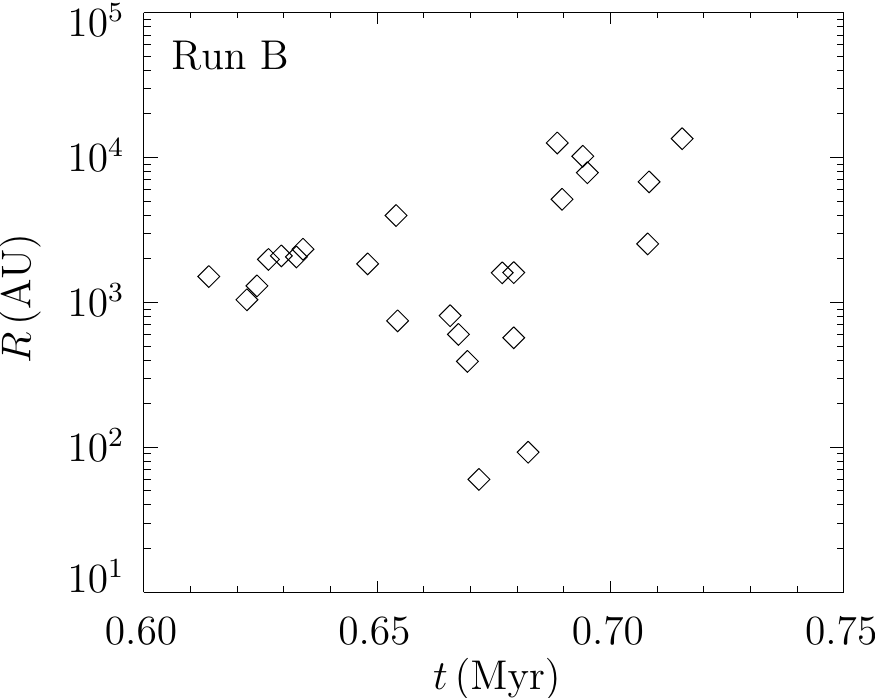}}
\centerline{\includegraphics[width=8cm]{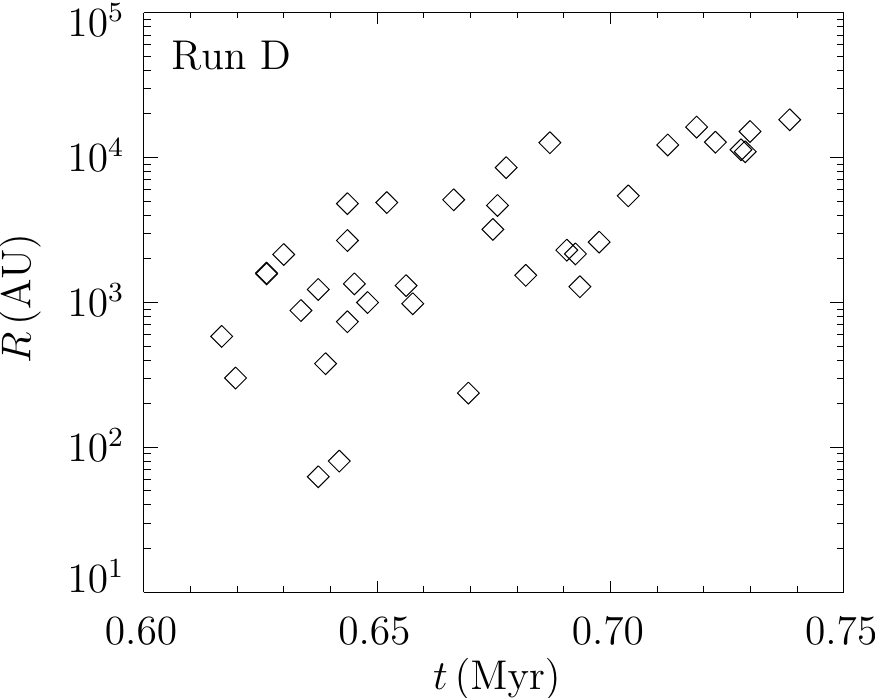}}
\caption{Radius of sink formation as
         a function of time
for run~B (\emph{top})
and run~D (\emph{bottom}). Because of the absence of radiation heating, both the Jeans
length and the Jeans mass are smaller in run~D than in run~B, giving rise to the formation
of more numerous, but generally lower-mass stars.
In both simulations, sink formation gradually occurs at larger disk radii. The accretion
heating by the first stars to form suppresses sink formation at small disk radii in run~B
until relatively late in the cluster evolution.}
\label{fig:radtim}
\end{figure}

To illustrate the tendency of star formation to occur at increasingly larger disk radii, we
show the disk radius at which new sink particles form as
a
function of time for
runs~B and D in Figure~\ref{fig:radtim}. Because the accretion heating sets in already with the very first
stars that form in run~B, sink formation is suppressed at small disk radii initially. The massive
stars slowly spiral outwards with time, so that at $t \approx 0.67\,$Myr their radiation can
         be
shielded by filaments in the disk. Within these filaments, the gas then cools until
local collapse sets in and two sink particles form near the center of the disk. Once the filament has
dissolved, the gas heats up again and no further sink particles form in the inner disk region.

\section{Discussion}
\label{sec:discussion}

\subsection{Initial Conditions}

\subsubsection{Density Profile}

As we argue above, the fragmentation behavior of the disk forming around the massive central star
depends sensitively on the initial and boundary conditions, i.\ e.\ on the physical properties of
the high-mass cloud core. One of the key parameters that determines whether fragmentation becomes
widespread during the collapse of a massive cloud core is its initial density profile. Numerical
simulations indicate that density profiles with flat inner core are more susceptible to
fragmentation, while centrally concentrated cores (for example such as singular isothermal spheres
with $\rho \propto r^{-2}$) usually form only one or at most a few objects \citep{girietal10}.

The density structure of prestellar cores is typically estimated through the analysis of dust emission
or absorption using near-IR extinction mapping of background starlight, mapping of millimeter/submillimeter
dust continuum emission, and  mapping of dust absorption against the bright mid-IR background emission
\citep{bertaf07}. A main characteristic of the density profiles derived with the above techniques is that
they require a central flattening. The density of low-mass cores is almost constant within radii smaller
than $2500 - 5000\,$AU with typical central densities of 10$^{5}$ -- 10$^{6}$ cm$^{-3}$
\citep{motetal98,wamoan99}.  A popular approach is to describe these cores as truncated isothermal
(Bonnor-Ebert) sphere \citep{ebert55,bonnor56}, that often  provides a good fit to the
data \citep{bacmanetal01,alvesetal01,kandorietal05}.
      Bonnor-Ebert spheres
are equilibrium solutions of self-gravitating gas
bounded by external pressure. However, such density structure is not
unique. Numerical calculations
of the dynamical evolution of supersonically turbulent clouds show that transient cores forming at the
stagnation points of convergent flows exhibit similar morphology \citep{baklva03,klessenetal05,banerjeeetal09}.
The situation is less clear when it comes to high-mass cores \citep{beutheretal07}, because most
high-mass cores studied to date show at least some sign of star formation or turn out to consist of
several sub-condensations when observed with high enough resolution
\cite[e.g.][]{beutetal05,beuthen09}. However, large-scale surveys, e.g. as conducted
in the Cygnus-X region \citep{motteetal07}, indicate that high-mass cores are in many aspects
similar to scaled-up versions of low-mass cores. We followed these lines of reasoning and chose an
initial density profile with flat inner core and $r^{-1.5}$ density profile outside a radius
of $r=0.5\,$pc.  
  Although we begin with smooth rather than turbulent initial
  conditions, by the time ionizing radiation begins to be emitted,
  gravitational fragmentation has already produced substantial
  density perturbations.

\subsubsection{Rotation}
The second important parameter is the initial rotation of this core, which defines the total amount of
angular momentum in the system and hence the radial extent of the protostellar accretion disk. To assess
the relevance of fragmentation-induced starvation for high-mass star formation we therefore must compare
our choice of $\beta = 0.05$ for the ratio between rotational and potential energy in our $1000\,M_\odot$
core both with observational data of molecular cloud cores and with numerical simulations of core formation.
         Our choice of
values correspond to a specific angular momentum of 
$j = 1.27 \times 10^{23}\,$cm$^2$s$^{-1}$.

The high-mass cloud cores observed by \citet{pirogovetal03} with masses up to a few thousand solar masses
show specific angular momenta $j$ from a few $\times 10^{22}\,$cm$^2$s$^{-1}$ up to $10^{23}\,$cm$^2$s$^{-1}$,
corresponding to $\beta$-values up to $0.07$. We took this sample as motivation for our choice of parameters.
Interesting in this context is also the  large-scale rotational motion of molecular gas around a cluster
of hypercompact \hii\ regions in G20.08-0.14~N seen by \citet{gavmadetal09}. This flow has a radius of about
$0.5\,$pc, roughly consistent with our simulation. Similar observations were made by \citet{keto90} for
G10.6-0.4 and \citet{welchetal87} for W94A.

It needs to be noted, however, that although the specific angular momenta of observed cores exhibit
considerable scatter there is a clear trend of decreasing $j$ with decreasing core mass $M$. For example, the
mean values in the sample discussed by \citet{goodmanetal93} are $j \approx 6 \times 10^{21}\,$cm$^2$s$^{-1}$
with masses up to several hundred solar masses, while \citet{casellietal02} studied low-mass cores with typically
only a few solar masses and find $j \approx 7 \times 10^{20}\,$cm$^2$s$^{-1}$. In all cases the typical values for
the ratio between rotational energy and potential energy is $\beta \lesssim 0.05$. This trend continues down
to stellar scales. A binary G star with a orbital period of 3 days has $j \approx 10^{19}\,$cm$^2$\,s$^{-1}$,
while the spin of a typical T~Tauri star is a few $\times 10^{17}\,$cm$^2$\,s$^{-1}$. Our own Sun rotates only
with $j\approx 10^{15}\,$cm$^2$\,s$^{-1}$. That means, during the process of star formation most of the initial
angular momentum is removed from the collapsed object \citep{boden95}. 

If we look for guidance from numerical simulations of molecular cloud formation and fragmentation
\citep{maclow04,klessenetal09} we see a very similar trend. \citet{gammieetal03} find a typical mean
specific angular momentum of $j = 4 \times 10^{22}\,$cm$^2$s$^{-1}$, while \citet{jappkles04} find values
in the range $10^{20}\,$cm$^2$s$^{-1} \lesssim j \lesssim10 ^{21}\,$cm$^2$s$^{-1}$ for low-mass cores in the
simulations by \citet{klessetal98} and \citet{klesburk01,klesburk00} depending on the
evolutionary state of protostellar collapse. See also \citet{tillpudr07} and \citet{offetal08}. Altogether,
we find that
our initial conditions are consistent with simulations of core formation.

\subsection{Relation to Other Theoretical Models}

\subsubsection{Monolithic Collapse Models}
\label{subsec:moncol}

The monolithic collapse model rests on the similarity between the shape of the observed core mass distribution
\citep{motetal98,johnstone00,johnstone06,ladaetal08} and the stellar initial mass function, IMF
\citep{kroupa02,chabrier03}. It assumes a one-to-one relation between the distributions with only a constant
efficiency factor separating the two functions. Each molecular cloud core collapses to form a single star or
at most a close binary system, with protostellar feedback processes determining the efficiency of the process
\citep{matzner00}. This model reduces the problem of the origin of the IMF to the problem of determining the
mass spectrum of bound cores, although strictly speaking the idea that the IMF is set by the mass spectrum of
cores is independent of any particular model for the origin of that mass spectrum. Arguments to explain the
core mass distribution generally rely on the statistical properties of turbulence
\citep{klessen01,padnor02,hencha08,hencha09}, which generate structures with a
pure powerlaw mass spectrum. The thermal Jeans mass in the cloud then imposes the flattening and turn-down
in the observed mass spectrum.

However, there are a number of caveats. Many of the prestellar cores found in observational surveys  appear to be
stable entities and thus are unlikely to be in a state of active star formation. In addition, the simple
interpretation that one core forms on average one star, and that all cores contain the same number of thermal
Jeans masses, leads to a timescale problem \citep{clark07} that requires differences in the core mass function
and the IMF. Other concern about this model comes from  hydrodynamic simulations, which  seem to indicate that
massive cores should fragment into many stars rather than collapsing monolithically
\citep{dobbsetal05,clabon06,bonbat06,federrathetal10}. This objection is, however, weakened by the fact that
magnetic fields are able to reduce the level of fragmentation on scales of molecular clouds as a whole
\citep{heitschetal01} as well as of collapsing cloud cores \citep{henfro08,hentey08}.
In addition, radiative feedback also is able to reduce the number of fragments that form during the collapse
of high-mass cores \citep{krumkleinmckee07}
as well as low-mass cores \citep{offetal09}.
But again, all current
simulations indicate that
neither accretion
heating nor ionizing radiation can prevent the fragmentation of the massive dense disk that builds up during
protostellar collapse, nor can they stop accretion onto the massive star \citep{yorke02,krumholzetal09,petersetal10a}.
     Instead, the heating merely increases the average mass of the fragments.
This is also supported by analytic estimates comparing typical accretion rates onto the disk with its ability
to transport matter inwards through viscous and gravitational torques \citep{krattmatz06,kratteretal10}.
Our current results
         agree
with these studies.

\subsubsection{Competitive Accretion}
\label{subsec:compaccr}

A second model for the origin of the IMF, called competitive accretion, focuses on the interaction between
protostars, and between a protostellar population and the gas cloud around it
\citep{bonnell01a,bonnell01b,bonbat02,batbon05}. In the competitive accretion picture the origin of the peak in
the IMF is similar to the monolithic collapse model, it is set by the Jeans mass in the prestellar gas
cloud. However, rather than fragmentation in the gas phase producing a spectrum of core masses, each of which
collapses down to a single star or star system, in the competitive accretion model all gas fragments down to
roughly the Jeans mass. Prompt fragmentation therefore creates a mass function that lacks the powerlaw tail at
high masses that we observe in the stellar mass function. This part of the distribution forms via a second
phase in which Jeans mass-protostars compete for gas in the center of a dense cluster. The cluster potential
channels mass towards the center, so stars that remain in the center grow to large masses, while those that are
ejected from the cluster center by $N$-body interactions remain low mass \citep{klesburk00,bonetal04}. In this
model, the apparent similarity between the core and stellar mass functions is an illusion, because the observed
cores do not correspond to gravitationally bound structures that will collapse to stars \citep{clabon06, smithetal08}.

The competitive accretion picture has been challenged, on the grounds that the kinematic structure observed in
star-forming regions is inconsistent with the idea that protostars have time to interact with one another strongly
before they completely accrete their parent cores \citep{krumetal05,andreetal07}.

Taken at face value, competitive accretion models show a correlation between the mass of the most massive
star $\mmax$ and the total cluster mass $\msinks$ during the whole cluster evolution
that is roughly $\mmax \propto \msinks^{2 / 3}$ \citep{bonetal04}. This correlation has
been argued to represent a way to observationally confirm competitive accretion \citep{krumbon07}
and is in fact in good agreement with observations \citep{weidkroup06,weidetal10}.
However, we find that our simulations also reproduce the observed relation between $\mmax$ and $\msinks$.  

\begin{figure}
\centerline{\includegraphics[width=7.5cm]{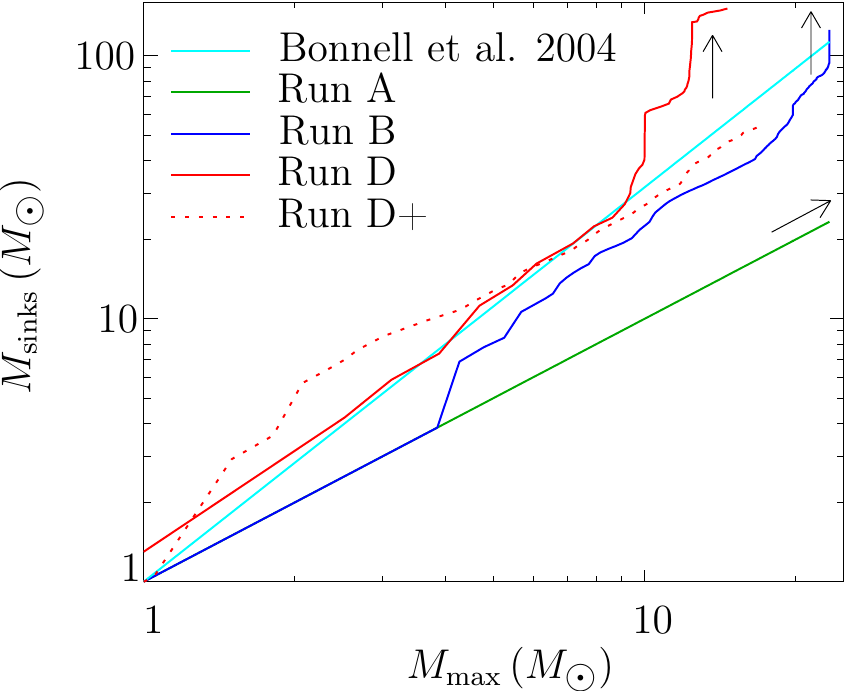}}
\caption{The total mass in sink particles $\msinks$ as a function of the most massive star in the
cluster $\mmax$. We plot the curves for run~A, run~B and run~D, the convergence study run~D+ as
well as the fit from the competitive accretion simulations \citet{bonetal04}. All simulations
follow the competitive accretion prediction with good
accuracy. The maximum mass $\mmax$ in run~B is always larger than in run~D for a fixed cluster mass
$\msinks$. The turn-off away from the scaling relation (indicated by arrows) is shifted towards higher
masses by radiative feedback. The higher resolution study run~D+ was not followed long enough
to show the turn-off.}
\label{fig:compaccr}
\end{figure}

Figure~\ref{fig:compaccr} shows $\msinks$ as function of $\mmax$ for run~A, run~B and run~D,
the higher resolution convergence study run~D+ and
the relation $\mmax = 0.39 \msinks^{2/3}$, which was found by \citet{bonetal04}
as a fit to their simulation data. Over the whole cluster evolution, the curve for run~D lies
above this fit, while the curve for run~B always lies below it. The
     fit agrees
with our simulation data
         as well as it does to that of
\citet{bonetal04}.
The higher resolution simulation run~D+ was only carried out for a
fraction of the total 
  time
of the other simulations, but its data also agrees very well with the fit, demonstrating convergence
of the results from the lower resolution simulations. This indicates that the
scaling is not unique to competitive accretion, but can also be found with the fragmentation-induced
starvation scenario and hence cannot be used as an observational confirmation of competitive
accretion models.

The agreement of both models
         to
this prediction is surprising because
the accretion behavior of the most massive stars is totally different. Whereas the massive stars
in competitive accretion simulations automatically accrete large fractions of the available gas
because they reside in the center of the gravitational potential during the whole cluster
evolution, the massive stars in fragmentation-induced starvation models have continuously
decreasing accretion rates since they are starved
         of
material by other cluster members and,
in the final phase, by feedback from ionizing radiation. 
It seems that the observed relation between  $\msinks$ and $\mmax$ is a very general result of protostellar
interaction in a common cluster environment, and not an unambiguous sign of competitive accretion at work. 

We can gain further insight from Figure~\ref{fig:compaccr}. It shows that the accretion heating suppresses
low-mass star formation and that for all times the simulation with feedback contains a more massive
star relative to the whole cluster mass than the simulation without feedback. When the most massive
star in run~D reaches $10\,$M$_\odot$, much more gas is used up to create additional low-mass stars
than is funneled towards the most massive one. Thus, the growth of the most massive star in run~D is
much more ineffective than in run~B. Remarkably, all four simulations in \citet{bonetal04} show
the turn-off towards accretion onto low-mass stars around $10\,$M$_\odot$, and only one simulation
has formed a star more massive than $10\,$M$_\odot$. A very similar turn-off (indicated by an arrow
in Figure~\ref{fig:compaccr}) at the same mass scale is
found in run~D, but the radiative heating in run~B shifts the turn-off towards higher masses, so
that the scaling is followed closely over a larger mass range. In fact, since the turn-off occurs
only at the end of the simulation and some of the massive stars are still accreting at this point,
it is unknown up to
          what
mass the scaling will continue.
        However,
these results clearly indicate that
radiative feedback is necessary to reproduce the scaling for masses
beyond $10\,$M$_\odot$
observed by \citet{weidetal10}.

\subsection{Relevance for Primordial Star Formation}

We note that the initial conditions adopted here may
         also be
appropriate for the formation of
metal-free Population III stars. The very first generation of stars in the universe is thought to
form at a redshift of $z \sim 15$ in relatively unperturbed and quiescent halos with masses of
$\sim 10^6\,$M$_\odot$ in dark matter and about $10^5\,$M$_\odot$ in baryons
\citep{bromlars04,brommetal09}. This gas has primordial composition, i.e. basically
consists of hydrogen and helium, and cools mostly via H$_2$ line emission
\citep[e.g.][]{annietal97,glover05,glovjapp07}. These halos have a well-defined
density structure and rotational profile \citep{abeletal02} which leads to the build
up of an extended accretion disk around the central object
(see, e.g., \citealt{yoshidaetal08} for numerical simulations, or \citealt{tanmckee04}
for
         an
analytical model). Just as discussed above, if the mass flow onto the disk exceeds its
capacity for transporting material inward by viscous torques, the disk becomes susceptible to
gravitational instability and fragments. There are first indications that this indeed happened
during Pop III star formation from high-resolution numerical simulations that follow gravitational
collapse beyond the formation of the central hydrostatic protostar
\citep{clarketal08,turketal09,stacyetal10}.
          We
propose that fragmentation-induced starvation may be the physical
process that determines the masses of Pop III stars.

          The situation
is very different in atomic cooling halos \citep{wiseabel07,greifetal09}.
Once the virial mass of a halo exceeds a value of $\sim 10^8\,$M$_\odot$, the infalling gas gets
partially ionized in the virialization shock. The enhanced abundance of free electrons triggers
rapid H$_2$ formation and some fraction of the gas cools down very rapidly. As a result there
are streams of cold gas that can travel all the way to the center \cite[see also][]{dekeletal09}
and the overall velocity structure of the gas becomes highly chaotic and turbulent. We expect
that these environmental conditions prevent the
build-up of a large-scale accretion disk in
the center of the halo and our model of fragmentation-induced starvation should therefore not
be applicable. However, a more thorough analysis awaits the execution of detailed,
high-resolution numerical simulations.

\section{Summary and Conclusion}
\label{sec:summary}

We have presented a detailed analysis of the fragmentation-induced starvation scenario, which
was first introduced by \citet{petersetal10a}. We have studied the accretion history and disk instability
in collapse simulations of massive star formation. We have compared the results from a full
simulation with radiation feedback and multiple sink particles with two control runs, one
with multiple sink particles but without radiation feedback and one with radiation feedback
but only a single sink particle. The combined analysis of all three simulations allows us
to establish fragmentation-induced starvation as a 
workable model of massive star formation.
We have also compared the new model to the monolithic collapse and competitive accretion
models and speculated on the relevance of fragmentation-induced starvation for primordial
star formation.

The basic principle of fragmentation-induced starvation is star formation in a
rotating
gravitationally unstable flow. Gravitational instability in such a flow leads to fragmentation
of the accretion flow and the formation of companions around the central massive star
          that
starve the central star
         of
infalling material. In our simulations, star formation occurs
within a rotationally flattened, disk-like structure. As more and more material falls onto
the disk, star formation proceeds radially outwards, keeping the total disk mass roughly
constant. The accretion history of individual stars is tightly connected with 
the position of these stars relative to the dense filaments and low-density voids in the disk.
The fact that most if not all high-mass stars are observed in higher-order multiple stellar systems
         that
themselves
         belong to
more massive star clusters provides strong evidence for the
widespread occurrence of fragmentation that forms the basis of our discussion. 

We find that accretion heating does not prevent fragmentation of the disk, but leads to
a higher local Jeans mass. As a result, fewer stars form with than without radiation feedback.
The accretion heating shifts the masses of the most massive stars up, but leaves
the average stellar mass almost unaffected. The accretion heating does not change the overall
star formation rate of the whole stellar cluster. Feedback by ionizing radiation is unable
to stop protostellar growth if the accretion flow is strong enough. However, if the accretion
flow weakens due to starvation, an \hii\ region can expand and terminate the accretion process. The growing
\hii\ regions reduce the total star formation rate and do not trigger star formation at the
ionization front.

Our model is able to explain the observed morphologies \citep{petersetal10b} and time variability
\citep{galvmadetal10inprep} of ultracompact \hii\ regions. We find that we can consistently reproduce
the observed relation between the total mass of the star cluster $\msinks$ and the maximum stellar mass
in the cluster $\mmax$, $\mmax \propto \msinks^{2 / 3}$. This relation seems to be the general outcome
of protostellar interaction in a common cluster environment rather than being a signpost of competitive
accretion only, as previously claimed. In fact, the dynamical processes discussed here exhibit exactly
the opposite behavior of competitive accretion, rather than run-away accretion onto the most massive
star together with the suppression of the growth of lower-mass objects, we see that angular momentum
conservation and the presence of lower-mass objects limit the mass growth of massive stars.   

Our simulations provide evidence for the rejection of proposals that the observed maximum stellar
mass of $\sim 100\,$M$_\odot$ is set by radiative feedback. When disk fragmentation is artificially
suppressed (run~A) the central protostar accretes material at very high rate unimpeded by the intense
UV radiation it emits without any indications of an upper limit \cite[see also][]{petersetal10a}. When
we permit disk fragmentation to occur, it is the process of fragmentation-induced starvation that
prevents the stellar mass to become larger than $\sim 25\,$M$_\odot$ with our choice of initial
conditions. We expect more massive, more centrally-condensed, and/or slower rotating  cloud cores
to lead to more massive protostars. Indeed, extensive parameter studies \citep{girietal10} show that
the initial density profile dominates the accretion behavior, explaining the formation of a
$40\,$M$_\odot$ star from a $100\,$M$_\odot$ core in \citet{krumholzetal09}. The
alternative view is to attribute the apparent stellar mass limit to internal stability constraints.

\acknowledgements{
T.P. is a Fellow of the {\em Baden-W\"{u}rttemberg Stiftung} funded by their program International
Collaboration II (grant P-LS-SPII/18). He also acknowledges support from an Annette Kade Fellowship for his
visit to the American Museum of Natural History and a Visiting Scientist Award of the Smithsonian
Astrophysical Observatory (SAO).
R.S.K.\ acknowledges financial support from the {\em Baden-W\"{u}rttemberg Stiftung}
via their program International Collaboration II (grant P-LS-SPII/18) and from the German
{\em Bundesministerium f\"{u}r Bildung und Forschung} via the ASTRONET project STAR FORMAT (grant 05A09VHA).
R.S.K. furthermore
gives
thanks for subsidies from the {\em Deutsche Forschungsgemeinschaft} (DFG) under
grants no.\ KL 1358/1, KL 1358/4, KL 1359/5, KL 1358/10, and KL 1358/11, as well as from a Frontier
grant of Heidelberg University sponsored by the German Excellence Initiative. R.S.K. also thanks the
KIPAC at Stanford University and the Department of Astronomy and Astrophysics at the University of
California at Santa Cruz for their warm hospitality during a sabbatical stay in spring 2010.
M.-M.M.L. was partly supported by NSF grant AST 08-35734.
R.B. is funded by the DFG via the Emmy-Noether grant BA 3607/1-1.
We acknowledge computing time at the Leibniz-Rechenzentrum in Garching (Germany), the NSF-supported
Texas Advanced Computing Center (USA), and at J\"ulich Supercomputing Centre (Germany). The FLASH code
was in part developed by the DOE-supported Alliances Center for Astrophysical Thermonuclear
Flashes (ASCI) at the University of Chicago. This work was supported in part by the U.S. Department
of Energy contract no. DE-AC-02-76SF00515. We thank the anonymous referee for very useful
comments.}

\end{document}